\documentclass[preprint,preprintnumbers,amsmath,amssymb]{revtex4}

\usepackage{graphicx}
\usepackage{dcolumn}
\usepackage{bm}
\begin{document}

\title{Can the quantum vacuum be used as a reaction medium to generate thrust?}

\author{T. Lafleur$^{1, 2}$}
\email{trevor.lafleur@lpp.polytechnique.fr}
\affiliation{$^{1}$ Laboratoire de Physique des Plasmas, CNRS, Sorbonne Universit\'{e}s, UPMC Univ Paris 06, Univ Paris-Sud, Ecole Polytechnique, 91128 Palaiseau, France}
\affiliation{$^{2}$ ONERA-The French Aerospace Lab, 91120 Palaiseau, France}

\begin{abstract}
A number of recent controversial experiments have observed anomalous thrust forces with devices making use of electromagnetic fields, and which do not appear to emit any particles or radiation; in apparent violation of the principle of the conservation of momentum.  This has led to proposals that the measured forces are being produced due to an interaction between the applied electromagnetic fields of the device, and virtual particles/anti-particles of the quantum vacuum.  Using an argument based on Lorentz invariance, as well as a formal quantum field theoretic calculation, we investigate the feasibility of these proposals.  We find that both the vacuum expectation value of the linear momentum operator (the zero-point momentum), as well as the conserved current, are both exactly zero; the quantum vacuum does not appear to be a viable reaction medium with which to generate thrust.         
\end{abstract}

\maketitle

\section{Introduction}
   
The quantum vacuum is well known to produce many striking and unexpected phenomena, including pair production, the Casimir effect, and vacuum polarization \cite{milonni, schweber, itzykson}.   These phenomena have no classical analogues, and are vital in understanding a number of physically measurable effects, such as the difference in the $^2$S$_{1/2}$ and $^2$P$_{1/2}$ energy levels of the hydrogen atom (the Lamb shift), and the anomalous magnetic dipole moment of the electron \cite{milonni}.  Pair production is also thought to play an important role in a number of cosmological processes (such as black hole decay \cite{hawking1, hawking2}), as well as in some heavy ion collisions \cite{pp1, pp2}.  Application of a very high strength electric field can result in vacuum breakdown where particle/anti-particle pairs are physically created from the vacuum (hence the term pair production).  While the critical electric field strengths needed to initiate this vacuum breakdown have not yet been achieved experimentally, high powered laser experiments are expected to reach the needed levels in the near future \cite{pp6}.  Nevertheless, pair production has a strong theoretical foundation in quantum field theory (QFT), and was first predicted by Schwinger more than 60 years ago \cite{schwinger}.  Since then, this phenomena has been studied extensively by many authors, with generalizations made to include different external electric and magnetic field configurations, as well as spatial and temporal variations of these configurations \cite{pp3, pp4, pp5, pp6, pp7, pp8, pp9, pp10, pp11, pp12, pp13}.

Another important phenomenon that has become synonymous with the quantum vacuum is the Casimir effect \cite{milonni}.  Here two parallel plates placed in very close proximity are observed to experience an attractive force.  This effect, first predicted by Casimir \cite{casimir}, has been verified in a number of independent experiments \cite{cas6, cas5} (often to a very high precision).  Like pair production, the Casimir effect has also been well studied theoretically, both for electromagnetic and massive (scalar and spinor) quantum fields \cite{milonni}, cylindrical and spherical geometries \cite{cas1, cas2, cas3, cas4, cas8}, and vacuums in the presence of external applied electromagnetic fields \cite{cas9, cas10}.  The standard explanation \cite{milonni} for this phenomenon is that only quantum vacuum modes that satisfy the boundary conditions imposed by the presence of the two plates can exist in the region between the plates, whereas outside of this region, all modes can exist.  This leads to a reduction in the zero-point energy of the vacuum between the plates compared with that outside, which results in an inwards pressure force exerted on the plates.  The Casimir effect is special in that it demonstrates that energy can be extracted from the vacuum (the inwards motion of the plates can be made to do work).  However, the plates move in opposite directions, so no net momentum is extracted.      

The extraction of a net momentum from the vacuum has been proposed by Feigel \cite{feigel} as a new quantum phenomenon that contributes to the momentum of dielectric media.  Here a dielectric material, in the presence of crossed external electric and magnetic fields, is observed to undergo motion due to momentum transfer from high frequency vacuum modes.  In such a situation the counter-propagating vacuum modes no longer eliminate each other (as they usually do in non-interacting quantum fields), and the vacuum fields gain a finite momentum which is compensated for by the opposite motion of the material.  For realistic dielectric materials, the effect has however been predicated to very small (producing material velocities of the order of $50$~nm/s), even in the presence of high strength electric ($10^5$~V/m) and magnetic fields ($17$~T).  The Feigel phenomenon has so far not been verified experimentally, and remains controversial with a number of theoretical points disputed \cite{feigel_comment1, feigel_comment2, feigel_reply1, feigel_reply2}.  Further work by van Tiggelen \textit{et al} \cite{vanTigg} has argued that the result of Feigel is not invariant under a Lorentz transformation, and predicts no momentum transfer in homogenous materials when Lorentz invariance is addressed.  A small momentum transfer is however predicated for a squeezed vacuum (that is, for a vacuum located between two parallel plates similar to the Casimir geometry) in the presence of crossed electric and magnetic fields.  This effect is calculated to be immeasurably small though (producing material velocities of the order of $10^{-17}$~nm/s), and has also been challenged theoretically \cite{vanTigg_comment, vanTigg_reply}.         

A different mechanism to extract momentum from the quantum vacuum has recently been postulated \cite{white_2, white_recent} to explain the results of a number of controversial thruster experiments \cite{march, china_emdrive, white_2, white_recent}.  These experiments typically consist of a device using some electromagnetic field configuration, and which does not apparently emit any particles or radiation.  If correct, such propellantless thrusters would violate the principle of the conservation of momentum.  It has been proposed that the results might be explained as an interaction of the thruster electric and magnetic fields with virtual particles of the quantum vacuum to generate a net thrust force \cite{white_2, white_recent}.  Here virtual particles/anti-particles are viewed as experiencing a net drift in the ${\bf E}\times{\bf B}$ direction, where, ${\bf E}$ and ${\bf B}$ are the applied electric and magnetic fields respectively.  These experiments still remain unverified however, and so far no rigorous theory supporting (or disputing) the justification of these proposals has been presented.  They nevertheless pose an interesting question: can the quantum vacuum be used as a reaction medium?  In this paper we perform a field theoretic calculation of a quantum field interacting with external electric and magnetic fields, and explicitly investigate the possibility of using the quantum vacuum as a reaction medium with which to generate a net thrust force as these recent proposals have suggested. 
   
\section{Theoretical model}
     
\begin{figure}
\includegraphics{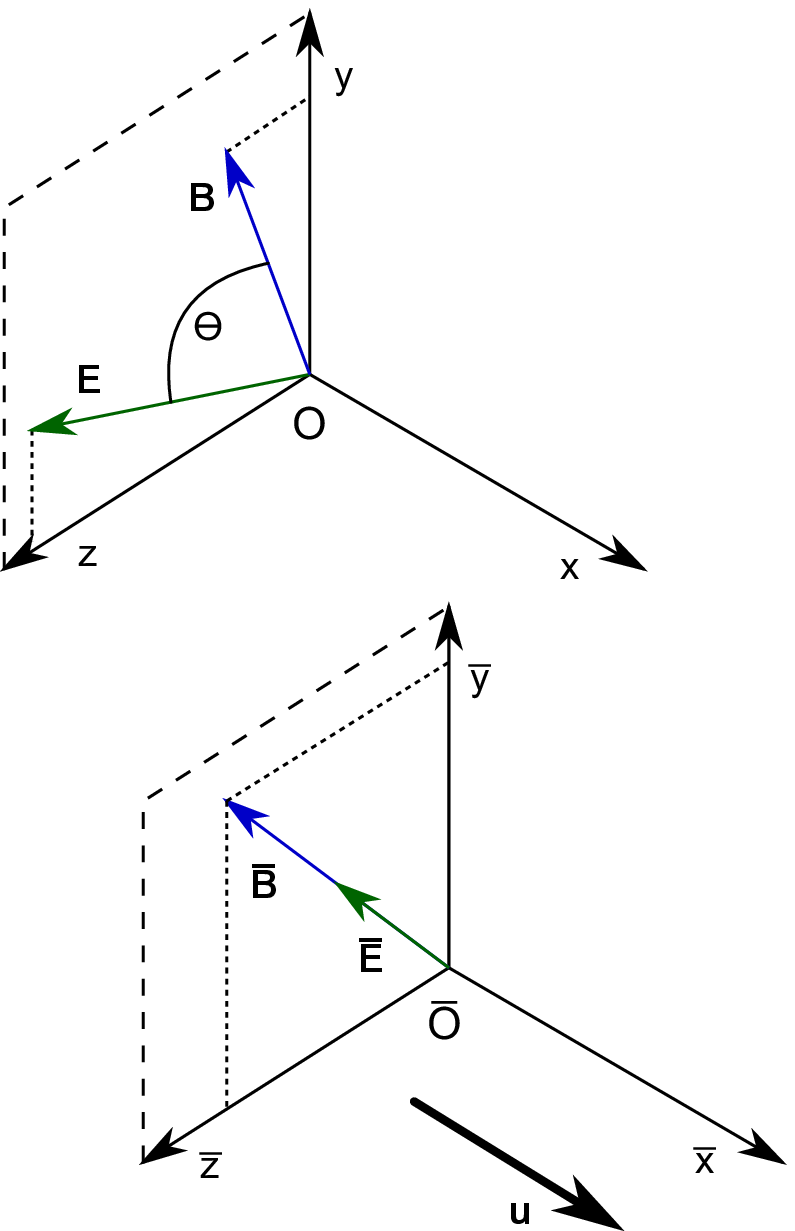}
\caption{}
\label{model}
\end{figure}

Consider a system with uniform, homogenous, electric, ${\bf E}$, and magnetic, ${\bf B}$, fields.  Without loss of generality, we can define a plane in reference frame $O$ (call it the $y$-$z$ plane) in which the electric and magnetic fields lie, as illustrated in Fig. \ref{model}.  The $x$-axis is then defined as the direction perpendicular to this plane.  The angle between the applied fields in the $y$-$z$ plane is $\theta$.  We assume here that the characteristic dimensions of any boundary structure of the system are sufficiently large and can be essentially ignored.  A physical test particle placed within the electric and magnetic field configuration of Fig. \ref{model} is well known to undergo a drift velocity, ${\bf v}$, given by
\newline
\begin{equation}
	{\bf v} = \frac{{\bf E}\times{\bf B}}{B^2}
	\label{drift1}
\end{equation}
\noindent   
\newline
where $B$ is the magnitude of the magnetic field.  Of particular relevance is the fact that this drift velocity is in the direction perpendicular to the applied fields, and is independent of the sign of the particle charge.  Thus particles of equal mass but opposite charge (such as electrons and positrons), will drift in the same direction.  The hypothesis \cite{white_2, white_recent} used to explain the anomalous thrust forces in Refs. \cite{march,china_emdrive, white_2, white_recent} postulates that a similar effect should occur with virtual particle/anti-particle pairs of the quantum vacuum, which is viewed as an infinite propellant supply.  Consequently, to test this hypothesis, we will focus on the field configuration given in Fig. \ref{model}.
       
Before proceeding further, it is worth noting that the presence of an applied electric field allows for the possibility of pair production from the vacuum.  This pair production rate has been calculated by a number of different authors, and the production rate \cite{milonni} (per unit volume and time) of spin-$1/2$ charged pairs, $w$, is 
\newline
\begin{equation}
	w = \frac{q^2E^2}{\pi^2\hbar^2c}\sum_{k = 1}^{\infty}\frac{1}{k^2}\exp\left(-\frac{k\pi m^2c^3}{\hbar qE}\right)
	\label{pairP1}
\end{equation}
\noindent   
\newline
where $q$ is the magnitude of the elementary charge.  The dominant term in this infinite series occurs for $k = 1$, which gives
\newline
\begin{equation}
	w \approx \frac{q^2E^2}{\pi^2\hbar^2c}\exp\left(-\frac{\pi m^2c^3}{\hbar qE}\right)
	\label{pairP2}
\end{equation}
\noindent   
\newline
Since most experimental electric field values have $E < 10^5$~V/m (and definitely those used in Refs. \cite{march, china_emdrive, white_2, white_recent}), we find that $w$ is almost exactly equal to zero.  Thus pair production is completely irrelevant to the experiments in Refs. \cite{march, china_emdrive, white_2, white_recent}, and the vacuum state persists. 

In passing, one might wonder whether pair production itself could be a viable means of continuously producing propellant for a future spacecraft, especially since the critical electric field strengths needed to achieve vacuum breakdown are expected to be achieved in a few years \cite{pp6}.  For a thruster operating at steady state, and continuously producing a mass rate $\dot{m}$ of particles, and $\dot{m}$ of anti-particles, the minimum required power input (assuming an ideal power source) to both produce the particles/anti-particles, and then accelerate them to some exhaust speed, $v$, would be 
\newline
\begin{equation}
	P = \frac{2\dot{m}c^2}{\sqrt{1 - v^2/c^2}} \approx 2\dot{m}c^2 + \dot{m}v^2
	\label{power}
\end{equation}
\noindent   
\newline             
Here for sake of argument we have assumed that the particle pairs are created at rest and then accelerated using some standard scheme (such as a series of biased electrodes for example) to their final exhaust speed.  Thus the approximation used in Eqn. \ref{power} applies, since usually $v/c << 1$.  The thrust generated by expelling the created particles is then, $F = 2\dot{m}v$, and so the thrust-to-power ratio of such an ideal pair-production thruster is
\newline
\begin{equation}
	\left(\frac{F}{P}\right)_{pp} = \frac{2v}{2c^2 + v^2} \approx \frac{v}{c^2}
	\label{ratio1}
\end{equation}
\noindent   
\newline
By comparison, the thrust-to-power ratio of an ideal photon thruster \cite{stuhlinger, griffiths} (that is, a propellantless thruster that emits photons/electromagnetic radiation) is equal to
\newline
\begin{equation}
	\left(\frac{F}{P}\right)_{photon} = \frac{1}{c}
	\label{ratio2}
\end{equation}
\noindent   
\newline           
Since $v/c << 1$, the photon thruster will always have a higher thrust-to-power efficiency than a pair production thruster.  Photon thrusters have also been technically demonstrated \cite{photon}, and are based on arguably simpler, and more well understood physics.  In addition, unless power is beamed to a pair production thruster (or unless solar panels are used to produce the required onboard power), any power source will need to supply the energy to create the particles, and since some percentage of mass is converted into energy in such systems (which is then reconverted back into the mass of the created particles/anti-particles), the effective propellant is in fact the power source itself.  The relevance of the above discussion to the present analysis is that pair production represents a sink for input power, and thus should be avoided.  Therefore the vacuum state needs to persist, and applied field strengths should be below the threshold values needed for vacuum breakdown.  
          
\section{Lorentz transformation}

For the QFT calculation to follow in the next section, the applied electric and magnetic fields in Fig. \ref{model} can more easily be treated if we perform a Lorentz transformation to a different reference frame \cite{dunne}, $\bar{O}$, moving at a speed $u$ (in the x-direction) relative to frame $O$, as illustrated in Fig. \ref{model}.  By choosing the speed of frame $\bar{O}$ correctly, the electric and magnetic fields will appear parallel (or anti-parallel) in this frame (so long as $\theta \neq \pi/2$ or $3\pi/2$).  In general, the electromagnetic field components, $F^{\lambda\sigma}$, in reference frame $O$, transform into field components, $\bar{F}^{\mu\nu}$, in $\bar{O}$ according to
\newline
\begin{equation}
	\bar{F}^{\mu\nu} = \Lambda_{\lambda}^{\mu}\Lambda_{\sigma}^{\nu}F^{\lambda\sigma}
	\label{lorentz1}
\end{equation}
\noindent   
\newline   
where $\Lambda_{\lambda}^{\mu}$ is the Lorentz transformation matrix.  If we align the co-ordinate axes of frames $O$ and $\bar{O}$, then the electromagnetic field components parallel to the motion of $\bar{O}$ remain unchanged (and in this case $\bar{E}_x = E_x = 0$ and $\bar{B}_x = B_x = 0$), while the transverse components become \cite{griffiths}
\newline
\begin{equation}
	\bar{E}_y = \gamma\left(E_y - uB_z\right)
	\label{lorentz2}
\end{equation}
\newline
\begin{equation}
	\bar{E}_z = \gamma\left(E_z + uB_y\right)
	\label{lorentz3}
\end{equation}
\newline
\begin{equation}
	\bar{B}_y = \gamma\left(B_y + \frac{u}{c^2}E_z\right)
	\label{lorentz4}
\end{equation}
\newline
\begin{equation}
	\bar{B}_z = \gamma\left(B_z - \frac{u}{c^2}E_y\right)
	\label{lorentz5}
\end{equation}
\noindent   
\newline
where $\gamma = \frac{1}{\sqrt{1 - u^2/c^2}}$, and barred quantities refer to the fields in frame $\bar{O}$, while unbarred quantities refer to the fields in frame $O$.  Requiring that the electric and magnetic fields be parallel in $\bar{O}$ then implies that
\newline
\begin{equation}
	\frac{\bar{E}_y}{\bar{E}_z} = \frac{\bar{B}_y}{\bar{B}_z}
	\label{lorentz6}
\end{equation}
\noindent   
\newline
which can be simplified using Eqns. \ref{lorentz2}-\ref{lorentz5} to give
\newline
\begin{equation}
	\frac{u}{1 + u^2/c^2} = \frac{E_yB_z - E_zB_y}{B^2 + E^2/c^2}
	\label{lorentz7}
\end{equation}
\noindent   
\newline
But ${\bf E}\times{\bf B} = (E_yB_z - E_zB_y)\hat{i}$, and thus in general we have
\newline
\begin{equation}
	\frac{{\bf u}}{1 + u^2/c^2} = \frac{{\bf E}\times{\bf B}}{B^2 + E^2/c^2}
	\label{lorentz8}
\end{equation}
\noindent   
\newline
Therefore, for given electric and magnetic fields in $O$, we can find the velocity of $\bar{O}$ needed to give parallel fields in this frame.
 
We now, without loss of generality, align the $y$-axis of frame $\bar{O}$ with the co-linear electric and magnetic fields.  It is at this point that we encounter the first indication of a null answer to the question posed in the introduction.  Because there is no preferred reference frame for the quantum vacuum (nor does it even have a rest frame), any force observed to be produced in $O$, must also be observed in $\bar{O}$.  However, relative to $\bar{O}$, there is a symmetry about the parallel fields, and thus no net force is expected.  The only way that such an observation can be reconciled with any apparent force observations made in $O$, is if the net force in $O$ is also zero.  In the next section we formally confirm this using a QFT calculation. 

\section{Charged scalar field}   
\subsection{Calculation of the quantum field}  

As an example QFT calculation, we consider a charged scalar field interacting with a classical external electromagnetic potential.  We set $\hbar$ and $c$ equal to $1$, assume a flat space-time with metric signature $(1, -1, -1, -1)$, and choose a gauge with a 4-potential given by $A^{\mu} = (0, -Bz, -Et, 0)$.  The electric and magnetic fields (we now drop the barred notation for convenience) are then
\newline
\begin{equation}
	{\bf E} = -\nabla V - \frac{\partial {\bf A}}{\partial t} = E\hat{{\bf j}}
	\label{gauge1}
\end{equation}
\newline
\begin{equation}
	{\bf B} = \nabla\times{\bf A} = B\hat{{\bf j}}
	\label{gauge2}
\end{equation}
\noindent   
\newline   
The Klein-Gordon equation, coupled to a electromagnetic potential, is
\newline
\begin{equation}
	\left(\partial^{\mu} - iqA^{\mu}\right)\left(\partial_{\mu} - iqA_{\mu}\right)\psi  + m^2\psi = 0
	\label{KG1}
\end{equation}
\noindent   
\newline
where $\psi = \psi(t, {\bf x})$ (and $\psi^\dagger = \psi^\dagger(t, {\bf x})$) is the scalar quantum field.  With the particular co-linear electromagnetic field configuration, and the choice of gauge made above, Eqn. \ref{KG1} can be written as    
\newline
\begin{equation}
	\left[\frac{\partial^2}{\partial t^2} - \frac{\partial^2}{\partial x^2} - \frac{\partial^2}{\partial y^2} - \frac{\partial^2}{\partial z^2} - 2iq\left(Bz\frac{\partial}{\partial x} + Et\frac{\partial}{\partial y}\right) + q^2\left(B^2z^2 + E^2t^2\right) + m^2\right]\psi = 0
	\label{KG2}
\end{equation}
\noindent   
\newline
We then perform a separation of variables, assuming solutions of the form, $\psi(t, {\bf x}) = \phi(t)\chi(z)e^{i\left(k_xx + k_yy\right)}$, which reduces Eqn. \ref{KG2} into the two separated equations  
\newline
\begin{equation}
	\left[\frac{d^2}{dt^2} + \left(k_y + qEt\right)^2 + m^2 + \beta\right]\phi = 0
	\label{KG3}
\end{equation}
\newline
\begin{equation}
	\left[\frac{d^2}{dz^2} - \left(k_x + qBz\right)^2 + \beta\right]\chi = 0
	\label{KG4}
\end{equation}
\noindent   
\newline
where $\beta$ is a constant.  These equations can be written in the form
\newline
\begin{equation}
	\frac{d^2\phi}{d\tau^2} + \left(\frac{1}{4}\tau^2 - a\right)\phi = 0
	\label{KG5}
\end{equation}
\newline
\begin{equation}
	\frac{d^2\chi}{d\xi^2} - \left(\xi^2 + K\right)\chi = 0
	\label{KG6}
\end{equation}
\noindent   
\newline  
where $\xi = \sqrt{\frac{1}{qB}}\left(k_x + qBz\right)$, $\tau = \sqrt{\frac{2}{qE}}\left(k_y + qEt\right)$, $K = \frac{\beta}{qB}$, and $a = -\frac{\left(m^2 + \beta\right)}{2qE}$.  Bounded solutions to Eqn. \ref{KG6} exist only for discrete values of $K$ (and therefore $\beta$), and are given by the associated Hermite polynomials \cite{griffiths_QM}  
\newline
\begin{equation}
	\chi(\xi)_n = \left(qB\right)^{1/4}\left(2^nn!\sqrt{\pi}\right)^{-1/2}e^{-\xi^2/2}H_n(\xi)
	\label{KG7}
\end{equation}
\noindent   
\newline
for $\beta_n = \left(2n + 1\right)qB$.  Two independent solutions ($\phi$ and $\phi^*$) to Eqn. \ref{KG5} are given in terms of the parabolic cylinder functions \cite{stegun}
\newline
\begin{equation}
	\phi_n = \frac{1}{\left(2qE\right)^{1/4}}e^{\frac{\pi a_n}{4} + \frac{i\pi}{2} + \frac{i\delta}{2}}D_{-ia_n - \frac{1}{2}}\left(\tau e^{-\frac{i\pi}{4}}\right)  
	\label{pcf1}
\end{equation}
\newline
\begin{equation}
	\phi^*_n = \frac{1}{\left(2qE\right)^{1/4}}e^{\frac{\pi a_n}{4} - \frac{i\pi}{2} - \frac{i\delta}{2}}D^*_{-ia_n - \frac{1}{2}}\left(\tau e^{-\frac{i\pi}{4}}\right) 
	\label{pcf2}
\end{equation}
\noindent   
\newline 
where $\delta = \textnormal{arg}$~$\Gamma\left(\frac{1}{2} + ia_n\right)$ and $a_n = -\frac{\left(m^2 + \beta_n\right)}{2qE}$.  We note in passing that the above results are similar to those obtained in Ref. \cite{pp9}, where the pair production rate in the presence of co-linear electric and magnetic fields was calculated. 

\subsection{Quantization of the quantum fields}

Using the separable solutions found in the section above, we now expand the fields in terms of annihilation and creation operators \cite{schweber} to give the general solution as 
\newline
\begin{equation}
	\psi = \sum_{n = 0}^{\infty}\int\frac{dk_xdk_y}{\left(2\pi\right)^2}\left[a_{{\bf k}}\chi_n \phi_n^* e^{i\left(k_xx + k_yy\right)} + b_{{\bf k}}^\dagger\chi_n \phi_ne^{i\left(k_xx + k_yy\right)}\right]
	\label{KG8}
\end{equation}
\noindent   
\newline 
and
\newline
\begin{equation}
	\psi^\dagger = \sum_{n = 0}^{\infty}\int\frac{dk_xdk_y}{\left(2\pi\right)^2}\left[a_{{\bf k}}^\dagger\chi_n \phi_n e^{-i\left(k_xx + k_yy\right)} + b_{{\bf k}}\chi_n \phi_n^* e^{-i\left(k_xx + k_yy\right)}\right]
	\label{KG9}
\end{equation}
\noindent   
\newline
where $a_{{\bf k}}$ and $a_{{\bf k}}^\dagger$ are annihilation and creation operators for particles, and $b_{{\bf k}}$ and $b_{{\bf k}}^\dagger$ are annihilation and creation operators for anti-particles.  Here the subscript ${\bf k}$ labels the states $(k_x, k_y, n)$.  We then have the vacuum states defined by $\left.\left\langle 0\right.\right|a_{{\bf k}}^\dagger = \left.a_{{\bf k}}\left|0\right\rangle\right. = 0$, and $\left.\left\langle 0\right.\right|b_{{\bf k}}^\dagger = \left.b_{{\bf k}}\left|0\right\rangle\right. = 0$.  For a charged scalar field interacting with an electromagnetic potential, the conjugate field momentum \cite{schweber} is  
\newline
\begin{equation}
	\pi\left(t, {\bf x}\right) = \dot{\psi}^\dagger\left(t, {\bf x}\right) + iqA_0\psi^\dagger\left(t, {\bf x}\right) = \dot{\psi}^\dagger\left(t, {\bf x}\right) 
	\label{mom1}
\end{equation}
\noindent   
\newline
where the second term is zero because of the choice of gauge (which has $A_0 = 0$).  To quantize the fields, we now impose the equal-time commutation relations
\newline
\begin{equation}
	\left[\psi\left(t, {\bf x}\right), \pi\left(t, {\bf x}'\right)\right] = i\delta^3\left({\bf x} - {\bf x}'\right)
	\label{mom2}
\end{equation}
\newline
\begin{equation}
	\left[\psi^\dagger\left(t, {\bf x}\right), \pi^\dagger\left(t, {\bf x}'\right)\right] = i\delta^3\left({\bf x} - {\bf x}'\right)
	\label{mom3}
\end{equation}
\noindent   
\newline
with all other commutators equal to zero.  We also postulate the following commutation relations for the annihilation and creation operators
\newline
\begin{equation}
	\left[a_{{\bf k}}, a_{{\bf k}'}^\dagger\right] = \left[b_{{\bf k}}, b_{{\bf k}'}^\dagger\right] = \left(2\pi\right)^2\delta_{nm}\delta\left(k_x - k_x'\right)\delta\left(k_y - k_y'\right)   
	\label{mom4}
\end{equation}
\noindent   
\newline
with all other commutation relations again being equal to zero.  Then, after simplifying, the left-hand side of Eqn. \ref{mom2} becomes
\newline
\begin{equation}
	\left[\psi, \dot{\psi}^\dagger\right] = \sum_{n = 0}^{\infty}\int\frac{dk_xdk_y}{\left(2\pi\right)^2}\chi_n\left(\xi\right)\chi_n\left(\xi'\right)e^{ik_x\left(x - x'\right) + ik_y\left(y - y'\right)}\left[\phi_n^*\frac{d\phi_n}{dt} - \phi_n\frac{d\phi_n^*}{dt}\right]  
	\label{mom5}
\end{equation}
\noindent   
\newline 
We notice that the term in square brackets is also equal to the definition of the Wronskian of the functions $\phi_n^*$ and $\phi_n$, from which we have the following identity \cite{stegun} for the parabolic cylinder functions
\newline
\begin{equation}
	W\left\{\phi_n\left(\tau\right), \phi_n^*\left(\tau\right)\right\} = \phi_n\frac{d\phi_n^*}{d\tau} - \phi_n^*\frac{d\phi_n}{d\tau} = -2i 
	\label{wronskian}
\end{equation}
\noindent   
\newline
Noting that $\frac{d}{d\tau} = \frac{d}{dt}\frac{dt}{d\tau}$, and that from Section IV A above we have $\frac{d\tau}{dt} = \sqrt{2qE}$, the term in square brackets in Eqn. \ref{mom5} reduces to $i$.  For the associated Hermite polynomials we make use of the completeness relation \cite{hermite}  
\newline
\begin{equation}
	\sum_{n = 0}^{\infty}\chi_n\left(\xi\right)\chi_n\left(\xi'\right) = \delta\left(\xi - \xi'\right)
	\label{hermite2}
\end{equation}
\noindent   
\newline
Again, from Section IV A, we remember that $\xi = \sqrt{\frac{1}{qB}}\left(k_x + qBz\right)$.  Thus $\xi - \xi' = \sqrt{qB}\left(z - z'\right)$.  Then using the identity \cite{ency}: $\delta\left(\alpha x\right) = \frac{\delta\left(x\right)}{\left|\alpha\right|}$, we have
\newline
\begin{equation}
	\sum_{n = 0}^{\infty}\chi_n\left(\xi\right)\chi_n\left(\xi'\right) = \delta\left(z - z'\right)
	\label{hermite3}
\end{equation}
\noindent   
\newline
Finally, using the well known identity \cite{griffiths_QM}
\newline
\begin{equation}
	\int\frac{dk_x}{2\pi}e^{ik_x\left(x - x'\right)} = \delta\left(x - x'\right)
	\label{delta1}
\end{equation}
\noindent   
\newline
Eqn. \ref{mom5} reduces to 
\newline
\begin{equation}
	\left[\psi, \dot{\psi}^\dagger\right] = i\delta^3\left({\bf x} - {\bf x}'\right)
	\label{total}
\end{equation}
\noindent   
\newline
as required.
 
\subsection{Linear momentum operator and conserved current}

With the quantum fields now known, we can begin to calculate some important vacuum expectation values to see if a zero-point momentum, or a net current exists.  The linear momentum operator is given by \cite{schweber}
\newline
\begin{equation}
	P^j = \dot{\psi}^\dagger\partial^j\psi + \dot{\psi}\partial^j\psi^\dagger
	\label{L_mom1}
\end{equation}
\noindent   
\newline  
and we are interested in the vacuum expectation value: $\left.\left.\left\langle 0\right|P^j\left|0\right\rangle\right.\right.$.  We begin with $P^x$.  Using Eqns. \ref{KG8}, \ref{KG9}, and the definition of the vacuum states, we can obtain
\newline
\begin{equation}
	\left.\left.\left\langle 0\right|P^x\left|0\right\rangle\right.\right. = \left.\left.\left\langle 0\right| \sum_{n = 0}^{\infty}\sum_{m = 0}^{\infty}\int\frac{dk_xdk_ydk_x'dk_y'}{\left(2\pi\right)^4}ik_x'\chi_n\chi_m\frac{d\phi_n^*}{dt}\phi_me^{-ik_xx - ik_yy + ik_x'x + ik_y'y}\left[b_{{\bf k}}b_{{\bf k}'}^\dagger - a_{{\bf k}}a_{{\bf k}'}^\dagger\right]\left|0\right\rangle\right.\right. 
	\label{L_mom2}
\end{equation}
\noindent   
\newline  
Then making use of Eqn. \ref{mom4} and simplifying, we obtain $\left.\left.\left\langle 0\right|P^x\left|0\right\rangle\right.\right. = 0$.  Similarly we find $\left.\left.\left\langle 0\right|P^y\left|0\right\rangle\right.\right. = 0$.  For $\left.\left.\left\langle 0\right|P^z\left|0\right\rangle\right.\right.$ the calculation is slightly more complicated, and after a similar procedure to that used above, we have
\newline
\begin{equation}
	\left.\left.\left\langle 0\right|P^z\left|0\right\rangle\right.\right. = 2\left.\left.\left\langle 0\right| \sum_{n = 0}^{\infty}\int\frac{dk_xdk_y}{\left(2\pi\right)^2}\chi_n\frac{d\chi_n}{dz}\frac{d\phi_n^*}{dt}\phi_n\left|0\right\rangle\right.\right. 
	\label{L_mom3}
\end{equation}
\noindent   
\newline  
Remembering that $k_x$ appears only in the terms $\frac{d\chi_n}{dz}$ and $\chi_n$, we focus now on the integral: $\int dk_x\frac{d\chi_n}{dz}\chi_n$.  Making the variable change, $\bar{k}_x = \sqrt{\frac{1}{qB}}\left(k_x + qBz\right)$, this integral can be written as 
\newline
\begin{equation}
	\int_{-\infty}^{\infty} dk_x\frac{d\chi_n}{dz}\chi_n = qB\int_{-\infty}^{\infty} \bar{dk}_x \frac{d\chi_n}{\bar{dk}_x}\chi_n = \frac{qB}{2}\int_{-\infty}^{\infty}\bar{dk}_x\frac{d}{\bar{dk}_x}\left(\chi_n\right)^2 = \frac{qB}{2}\int_{0}^{0} d\left(\chi_n\right)^2 = 0
	\label{L_mom4}
\end{equation}
\noindent   
\newline
Here the integration limits are zero in the last step because $\chi_n \rightarrow 0$ as $\bar{k}_x \rightarrow \infty$.  Thus $\left.\left.\left\langle 0\right|P^z\left|0\right\rangle\right.\right. = 0$, and therefore we have the result that the zero-point momentum of the field is exactly zero.   Consider now the conserved current \cite{schweber}, which is given by
\newline
\begin{equation}
	J^j = iq\left[\left(\partial^j\psi^\dagger\right)\psi - \psi^\dagger\partial^j\psi\right] - 2q^2A^j\psi^\dagger\psi
	\label{current1}
\end{equation}
\noindent   
\newline
Proceeding to calculate the vacuum expectation values using a similar procedure to that for the linear momentum operator, we easily find that $\left.\left.\left\langle 0\right|J^z\left|0\right\rangle\right.\right. = 0$.  The calculations for $\left.\left.\left\langle 0\right|J^x\left|0\right\rangle\right.\right. = 0$ and $\left.\left.\left\langle 0\right|J^y\left|0\right\rangle\right.\right. = 0$ are again slightly more complicated.  Consider the $\left.\left.\left\langle 0\right|J^x\left|0\right\rangle\right.\right. = 0$ calculation
\newline
\begin{equation}
	\left.\left.\left\langle 0\right|J^x\left|0\right\rangle\right.\right. = \left.\left.\left\langle 0\right|2q\sum_{n = 0}^\infty\int\frac{dk_xdk_y}{\left(2\pi\right)^2}\chi_n^2\phi_n\phi_n^*\left(k_x + qBz\right)\left|0\right\rangle\right.\right. 
	\label{current2}
\end{equation}
\noindent   
\newline
Defining $\bar{k}_x = \sqrt{\frac{1}{qB}}\left(k_x + qBz\right)$, we can write the $dk_x$ integral as
\newline
\begin{equation}
	\int_{-\infty}^{\infty} dk_x\chi_n^2\left(k_x + qBz\right) = qB\int_{-\infty}^{\infty} d\bar{k}_x\chi_n^2\bar{k}_x
\label{current3}
\end{equation}
\noindent   
\newline
However, because the integrand is asymmetric over the integration range, this integral is equal to $0$.  Thus we have $\left.\left.\left\langle 0\right|J^x\left|0\right\rangle\right.\right. = 0$.  Similarly (noting that $\phi_n\phi_n^* = |\phi_n|^2$), $\left.\left.\left\langle 0\right|J^y\left|0\right\rangle\right.\right. = 0$.  Thus the conserved current is also identically zero.  The above QFT analysis therefore confirms the zero net force argument given at the end of Section III.  

\section{Conclusions}
                    
In the analysis above we have investigated the possibility of using the quantum vacuum as a reaction medium with which to generate thrust for a future spacecraft.  This has primarily been motivated by recent hypotheses which attempt to explain the anomalous experimental force measurements in Refs. \cite{march, china_emdrive, white_recent, white_2} as an interaction with virtual particles of the vacuum.  However, from an argument based on Lorentz invariance, as well as a formal quantum field theory calculation, we have shown that the zero-point momentum and the conserved current of the field are zero.  The quantum vacuum does not appear to be a feasible medium with which to generate a net force or extract a net momentum with the configurations investigated.  In addition, based on a simple discussion comparing the performance of a pair production thruster to that of a photon thruster, we find that pair production is unlikely to be a useful mechanism to continuously generate propellant.    

For the analysis in the present work we have made use of a Lorentz transformation to simplify the resulting QFT equations.  This relies on the use of the assumption of homogeneous electric and magnetic fields.  If there is a spatial variation, then this argument cannot be used so simply, because in this case there is no single reference frame in which the fields now appear co-linear everywhere.  In the context of pair production, such situations have required numerical calculations to solve the resulting equations \cite{pp13}.  It seems unlikely though that spatial variations alone would produce a net force.  Although the extraction of a net momentum has been postulated in inhomogeneous vacuums \cite{vanTigg} (due to a different mechanism than that discussed in this paper), the effect was found to immeasurably small, and it remains unclear whether this small non-zero value is an artefact of the field regularization techniques used.  Regardless though, if a continuous net force is indeed being produced in the experiments in Refs. \cite{march, china_emdrive, white_2, white_recent}, one might expect that this should also imply an anomalously high power loss from the electric circuits of the thruster device.  This power loss would be in addition to any standard power losses associated with such things as: ohmic heating, eddy current losses, dielectric and ferrite heating, radiation losses, etc.  Consequently, a dedicated effort to isolate known power losses from the total input power of the device should identify any additional anomalous losses.
                                            
\section*{Acknowledgements}
The author would like to thank Robert de Mello Koch at The University of the Witwatersrand, and Razvan Gurau at Ecole Polytechnique, for a number of useful discussions.

\newpage

\section*{Figure captions}
\noindent
{\bf Fig.~1}:  Schematic showing the original reference frame $O$, with the electric, ${\bf E}$ and magnetic, ${\bf B}$, fields, as well as a second reference frame, $\bar{O}$ moving at a speed $u$ parallel to the $x$-axis.  In this reference frame the electric, $\bar{{\bf E}}$, and magnetic, $\bar{{\bf B}}$, fields appear co-linear.           
\newline
\newline
\noindent

\newpage
\section*{References}

\begin{thebibliography}{10}

\bibitem{milonni} P. Milonni, \textit{The quantum vacuum}, (Wiley, New Jersey, 2005).
\bibitem{schweber} S.S. Schweber, \textit{An introduction to relativistic quantum field theory}, (Harper \& Row, New York, 1962)
\bibitem{itzykson} C. Itzykson and J.B. Zuber, \textit{Quantum field theory}, (McGraw-Hill, New York, 1980)
\bibitem{hawking1} S.W. Hawking, Commun. math. Phys. {\bf 43}, 199 (1975).
\bibitem{hawking2} G.W. Gibbons and S.W. Hawking, Phys. Rev. D {\bf 15}, 2738 (1977).
\bibitem{pp1} G. Baur, Phys. Rev. A {\bf 42}, 5736 (1990).
\bibitem{pp2} D.L. Burke \textit{et al}, Phys. Rev. Lett. {\bf 79}, 1626 (1997).
\bibitem{pp6} A.R. Bell and J.G. Kirk, Phys. Rev. lett. {\bf 101}, 200403 (2008).
\bibitem{schwinger} J. Schwinger, Phys. Rev. {\bf 82}, 664 (1951).
\bibitem{pp3} S.P. Kim and D.N. Page, Phys. Rev. D {\bf 73}, 065020 (2006).
\bibitem{pp4} S.P. Kim and D.N. Page, Phys. Rev. D {\bf 65}, 105002 (2002).
\bibitem{pp5} C. Bottcher and M.R. Strayer, Phys. Rev. D {\bf 39}, 1330 (1989).
\bibitem{pp7} E. Brezin and C. Itzykson, Phys. Rev. D {\bf 2}, 1191 (1970).
\bibitem{pp8} J.K. Daugherty and I. Lerche, Phys. Rev. D {\bf 14}, 340 (1976).
\bibitem{pp9} N. Tanji, Ann. Phys. {\bf 324}, 1691 (2009).
\bibitem{pp10} H. Kleinert, R. Ruffini, and S.S. Xue, Phys. Rev. D {\bf 78}, 025011 (2008).
\bibitem{pp11} A. Blinne and H. Gies, Phys. Rev. D {\bf 89}, 085001 (2014).
\bibitem{pp12} Q. Su, W. Su, Q.Z. Lv, M. Jiang, X. Lu, Z.M. Sheng, and R. Grobe, Phys. Rev. Lett. {\bf 109}, 253202 (2012).
\bibitem{pp13} Q.Z. Lv, A.C. Su, M. Jiang, Y.J. Li, R. Grobe, and Q. Su, Phys. Rev. A {\bf 87}, 023416 (2013).
\bibitem{casimir} H.B.G. Casimir, Proc. Kon. Ned. Akad. Wetenschap. Ser. B {\bf 51}, 793 (1948).
\bibitem{cas5} S.K. Lamoreaux, Rep. Prog. Phys. {\bf 68}, 201 (2005).
\bibitem{cas6} S.K. Lamoreaux, Phys. Rev. Lett. {\bf 78}, 5 (1998).
\bibitem{cas1} M. Bordag, Phys. Rev. D {\bf 73}, 125018 (2006).
\bibitem{cas2} A. Rodriguez, M. Ibanescu, D. Iannuzzi, F. Capasso, J.D. Joannopoulos, and S.G. Johnson, Phys. Rev. Lett. {\bf 99}, 080401 (2007).
\bibitem{cas3} S.J. Rahi, T. Emig, R.L. Jaffe, and M. Kardar, Phys. Rev. A {\bf 78}, 012104 (2008).
\bibitem{cas4} F.C. Lombardo, F.D. Mazzitelli, and P.I. Villar, Phys. Rev. D {\bf 78}, 085009 (2008).
\bibitem{cas8} M. Bordag, E. Elizalde, K. Kirsten, and D. Leseduarte, Phys. Rev. D {\bf 56}, 4896 (1997).
\bibitem{cas9} M.V. Cougo-Pinto, C. Farina, M.R. Negrao, and A.C. Tort, J. Phys. {\bf A32}, 4457 (1999).
\bibitem{cas10} M. Ostrowski, Acta Phys. Polon. {\bf B37}, 1753 (2006).    
\bibitem{feigel} A. Feigel, Phys. Rev. Lett. {\bf 92}, 020404 (2004).
\bibitem{feigel_comment1} B.A. van Tiggelen and G.L.J.A. Rikken, Phys. Rev. Lett. {\bf 93}, 268903 (2004).
\bibitem{feigel_comment2} R. Sch\"{u}tzhold and G. Plunien, Phys. Rev. Lett. {\bf 93}, 268901 (2004).
\bibitem{feigel_reply1} A. Feigel, Phys. Rev. Lett. {\bf 93}, 268904 (2004).
\bibitem{feigel_reply2} A. Feigel, Phys. Rev. Lett. {\bf 93}, 268902 (2004).
\bibitem{vanTigg} B.A. van Tiggelen, G.L.J.A. Rikken, and V. Krsti\'{c}, Phys. Rev. Lett. {\bf 96}, 130402 (2006).
\bibitem{vanTigg_comment} G. Rousseaux, Phys. Rev. Lett. {\bf 100}, 248901 (2008).
\bibitem{vanTigg_reply} B.A. van Tiggelen and G.L.J.A. Rikken {\bf 100}, 248902 (2008).
\bibitem{white_2} H. White, P. March, N. Willians, and W. ONeil, \textit{Eagleworks laboratories: advanced propulsion physics research}, JANNAF Joint Propulsion Meeting, 5-9 December 2011, Huntsville, Alabama.
\bibitem{white_recent} D.A. Brady, H.G. White, P. March, J.T. Lawrence, and F.J. Davies, \textit{Anomalous thrust production from an RF test device measured on a low-thrust torsion pendulum}, 50th AIAA/ASME/SAE/ASEE Joint Propulsion Conference \& Exhibit, 28-30 July 2014, Cleveland, Ohio.
\bibitem{march} P. March and A. Palfreyman, AIP Conf. Proc. {\bf 813}, 1321 (2006).
\bibitem{china_emdrive} Y. Juan, W. Yu-Quan, M. Yan-Jie, L. Peng-Fei, Y. Le, W. Yang, and H. Guo-Qiang, Chin. Phys. B {\bf 22}, 050301 (2013).    
\bibitem{stuhlinger} E. Stuhlinger, \textit{Ion propulsion for space flight}, (McGraw-Hill, New York, 1964).
\bibitem{griffiths} D.J. Griffiths, \textit{Introduction to Electrodynamics}, Third Edition, (New Jersey, Prentice Hall, 1999).
\bibitem{photon} K. Toki, N. Asakura, T. Ohtsuka, and T. Akazawa, Experiments on Propellant-less Electric Propulsion Using Photon Pressure \textit{44th AIAA/ASME/SAE/ASEE Joint Propulsion Conference \& Exhibit}, Hartford, 21-23 July 2008.
\bibitem{dunne} G. Dunne, in \textit{From fields to strings}, edited by M. Shifman (World Scientific, Singapore, 2004), Vol. 1. 
\bibitem{griffiths_QM}  D.J. Griffiths, \textit{Introduction to quantum mechanics}, Second Edition, (New Jersey, Prentice Hall, 2004). 
\bibitem{stegun} M. Abramowitz and I.A. Stegun, \textit{Handbook of mathematical functions}, (Dover, New York, 1972).
\bibitem{hermite} N. Weiner, \textit{The fourier integral and certain of its applications}, (Cambridge Univeristy Press, Cambridge, 1988).
\bibitem{ency} M. Hazewinkel Ed., \textit{Encyclopedia of mathematics}, (Kluwer Academic Publishers, Dordrecht, 2001).

\end{thebibliography}
\end{document}